\documentclass[12pt]{article}
\usepackage[affil-it]{authblk}
\usepackage{amssymb, physics,slashed,graphicx,color}
\usepackage{hyperref} 
\usepackage{graphicx}
\usepackage{mathtools}
\begin{document}
\begin{titlepage}
\title{\bf Hint to Supersymmetry from GR Vacuum}	
\author{
\textsc{Gia Dvali$~^1$, Archil Kobakhidze$~^2$ and Otari Sakhelashvili$~^2$} 
\vspace{0.2cm} \\

\normalsize \itshape
$^1~$Arnold Sommerfeld Center, Ludwig-Maximilians-Universität,\\ \normalsize \itshape
Theresienstraße 37, 80333 München, Germany and \\ \normalsize \itshape
Max-Planck-Institut für Physik, Föhringer Ring 6, 80805 München, Germany \vspace{0.2cm} \\ 
\normalsize \itshape
$^{2}~$Sydney Consortium for Particle Physics and Cosmology, \\
\normalsize  \itshape
School of Physics, The University of Sydney, NSW 2006, Australia \\
}

\clearpage\maketitle
\thispagestyle{empty}
 
	\begin{abstract}
The $S$-matrix formulation of gravity suggests that the $\theta$-vacuum structure 
must not be sustained by the theory. We point out that, when applied to the vacuum of general relativity,  this criterion hints to supersymmetry.  
  The topological susceptibility 
 of gravitational vacuum induced by  Eguchi-Hanson instantons can be eliminated 
 neither by spin-$1/2$ fermions nor by an axion coupled via them 
 since such fermions do not provide instanton zero modes. Instead,  the job is done by a spin-$3/2$ fermion, hence realizing a  
 local supersymmetry.  
 This scenario also necessitates the spontaneous breaking of supersymmetry and predicts the existence of
 axion of $R$-symmetry which gets mass exclusively from 
 the gravitational instantons.  The $R$-axion can be a viable dark matter candidate.  Matching between the index and the anomaly imposes a constraint that spin-$1/2$ fermions should not contribute to the chiral gravitational anomaly.  

	\end{abstract}
 \pagebreak
\end{titlepage}
\section{Introduction}
Einstein's General Relativity (GR) is an extraordinarily successful theory of gravity.   It represents a theory 
 of a dynamical metric $g_{\mu\nu}(x)$ and utilizes the diffeomorphism invariance. The following action describes a minimal theory without any matter,
\begin{equation}
    S\, =\, \frac{M_{pl}^2}{2}\int d^4 x \sqrt{-g}\left(R(g)-2\Lambda\right),
\end{equation}
where $M_{pl}$ is the reduced Planck mass and $\Lambda$ is the cosmological term. 

 GR is a classical theory.  In this theory, any solution of Einstein's equations can be viewed as a legitimate background spacetime provided it satisfies standard energy conditions and the causality requirements. Any such solution can be treated as a vacuum for performing a weak-field expansion in small metric perturbations  $\delta g_{\mu\nu}(x)$ around it. In this sense, in classical theory there is no preferred background among the three maximally symmetric cosmological solutions of de Sitter, anti-de Sitter (AdS) and Minkowski,  corresponding to $\Lambda > 0,\, \Lambda <0$ and $\Lambda =0$ respectively. In particular, in classical theory all three unperturbed backgrounds are future-eternal. 
   
 In quantum theory the story changes. 
 A classical metric perturbation must be viewed 
 as an expectation value of a quantum field 
 over a proper coherent state,
 \begin{equation}
     \delta g_{\mu\nu}(x)  \, = \,  
     \frac{\langle \hat{h}_{\mu\nu}(x) \rangle}{M_{pl}} \,.
  \end{equation}    
  The operator $\hat{h}_{\mu\nu}(x)$ is the graviton 
  field which propagates massless quanta of spin$=2$. The term ``quantum gravity" shall be defined as a theory 
  that in its spectrum contains such a field. 
    A well-known representative is string theory in 
    which the graviton arises as the lowest vibration mode of a 
   closed string. 
   
   Quantum gravity being a more fundamental 
    theory is also more restrictive. In particular, not 
    every background of GR represents 
    a valid vacuum in quantum theory. 
     In fact, among the three maximally-symmetric 
     backgrounds,  Minkowski is singled out as the only 
     consistent vacuum with the possibility of non-trivial cosmological 
     history.  This is due to the following reasons.  
  
  First, it has been argued in ~\cite{Dvali:2013eja, Dvali:2014gua, Dvali:2017eba} that quantum gravity 
   excludes de Sitter as a valid vacuum. 
   The point is that in quantum theory the would-be classical de Sitter metric must be described as a coherent state 
   of gravitons.
   Of course, the consistency of such a state must be scrutinized by 
   the criteria such as BRST invariance, achieved by 
   an explicit construction ~\cite{Berezhiani:2021zst}.
     However, independently of specifics of the construction, already  
 from very general arguments it is evident that   
 de Sitter undergoes the phenomenon of 
   ``quantum breaking", manifested as a complete departure of the quantum evolution from the classical one.  
     This happens due to a
   back reaction from  Gibbons-Hawking radiation \cite{Gibbons:1977mu} which leads to a loss of coherence 
  by the coherent state of gravitons that describes a classical de Sitter metric ~\cite{Dvali:2013eja, Dvali:2014gua, Dvali:2017eba}. After quantum breaking, the expectation values can no longer match the solution of Einstein's equations with the classical source $\Lambda$.  Therefore, 
 quantum gravity excludes a positive cosmological constant as a viable source. 
   
   The quantum inconsistency of de Sitter can be understood  
 as a direct consequence of the $S$-matrix formulation of quantum gravity \cite{Dvali:2020etd}.  Currently, this is the only 
 existing formulation that is organic to string theory but goes beyond.  That is, in quantum gravity the $S$-matrix is not just a computational tool but represents the formulation of the theory. The requirement of a well-defined $S$-matrix imposes certain necessary conditions such as the existence of a globally-defined time as well as the existence of a rigid $S$-matrix vacuum.  None of these conditions can be satisfied in de Sitter \cite{Dvali:2020etd}. 
   
   In general, the $S$-matrix formulation of gravity excludes 
  any cosmological background that does not asymptote 
  to the $S$-matrix vacuum, e.g.,  such as the big crunch cosmologies. 
   This criterion eliminates universes with $\Lambda > 0$ along with any universe with $\Lambda < 0$  with a non-trivial cosmology. 
     Thus, the vacuum landscape of $S$-matrix gravity 
     represents a set of Minkowski vacua and 
   isolated AdS vacua without a possibility of cosmological time evolution. 
   
   It is also important for the consistency of the 
   $S$-matrix formulation that 
   Minkowski vacua are strictly stable, with no possibility of 
   decay via tunnelling \cite{Dvali:2011wk}. 
    The arguments leading to a similar conclusion of the impossibility of tunnelling from Minkowski were previously made by Zeldovich \cite{Zeldovich:1974py}.  
   According to the criterion of \cite{Dvali:2011wk} the energy splitting between the neighbouring Minkowski and AdS vacua 
   must be below the Coleman-De Luccia bound \cite{Coleman:1980aw}, which guarantees the absence of tunnelling \footnote{Interestingly, this provides a justification for the proposal of ``multiple point criticality" \cite{Bennett:1993pj, Bennett:1996vy} which 
   postulates that the vacua of the Standard Model are (nearly) degenerate. This proposal leads, in particular, to the prediction of the Higgs mass within the ballpark of the current experimental value \cite{Froggatt:1995rt}.}. 
   For instance,  supersymmetric AdS vacua are fully consistent.   
    This is also supported by the AdS/CFT correspondence  
    \cite{Maldacena:1997re}. 
       
The $S$-matrix constraint has important implications for the topological vacuum structure of the gauge theories and gravity. The existence of a topological structure of the vacuum, via corresponding instantons, results in the existence 
    of the landscape of $\theta$-vacua.   That is, the theory 
   generates a continuum of non-degenerate vacuum 
   states, belonging to different super-selection sectors. 
   No transitions among the different vacua are possible. Since their energies are different, even at the expense 
 of fine-tuning, only one of such vacua can be in Minkowski. 
  The rest, in general, will conflict with the $S$-matrix.
  
  Thus, the $S$-matrix gravity demands that 
  $\theta$-vacua in all sectors must be unphysical
  \cite{Dvali:2018dce, Dvali:2022fdv, Dvali:2023llt}. 
  In ordinary gauge theories, this requirement is translated 
  as the necessary presence of an axion or a chiral fermion
  with the global symmetry exclusively broken by the 
  instantons responsible for the given $\theta$-vacuum structure. 
  
In the present paper, we apply the  $S$-matrix 
  constraint to gravity demanding the absence of the 
 gravitational $\theta$-vacua.  That is, the criterion boils 
 down to the requirement that gravitational vacuum 
 must be $CP$-conserving. One can trade this requirement 
 as our starting point. With it, we arrive at the conclusion that in the minimal case, this necessitates the existence of a massless spin$=3/2$
  fermion. We interpret this result as an 
   \emph{evidence for supersymmetry}. 
  
  The outline of the argument is as follows. In pure Einsteinian gravity there exist Eguchi-Hanson instantons \cite{Eguchi:1978xp, Eguchi:1978gw} and their multi-instanton generalizations \cite{Gibbons:1978tef}.  Their action is regulated by the coefficient of the Gauss-Bonnet term, which must be non-zero for the validity of the effective field theoretic  (EFT) description of GR. With the Gauss-Bonnet term,  the action of 
 gravitational instantons is non-zero and they create the $\theta$-vacuum structure for gravity. In order to satisfy the $S$-matrix criterion, the $\theta$-vacua must be rendered unphysical. 
 
  This can be achieved neither by chiral spin-$1/2$ fermions 
  nor via an axion that couples to gravitational anomaly via 
  such fermions.  The reason is the absence of 
  zero modes for spin-$1/2$ fermion in the corresponding 
  gravitational instanton, as this is imposed by the index theorem
  \cite{Atiyah:1975jf}. Given this, the minimal working option is the existence of a spin-$3/2$ fermion. Such a fermion has zero modes in the instanton background \cite{ Eguchi:1978gw, Hanson:1978uv}. Correspondingly, it eliminates the gravitational $\theta$-vacua. That is, a  spin-$3/2$ fermion has the same effect on the gravitational $\theta$-vacuum as a chiral quark 
  has on the $\theta$-vacuum of QCD. The existence of an elementary 
  spin-$3/2$ fermion is an unambiguous signal of local supersymmetry.  The corresponding 
   chiral $U(1)$-symmetry that eliminates the gravitational 
   $\theta$-term is the $R$-symmetry. 
   
Thus, we arrive at the conclusion that the minimal extension of GR that satisfies the requirement of absence of $\theta$-vacua is supergravity. The consistency of the picture also 
demands that both $R$-symmetry and SUSY are broken spontaneously. 
Correspondingly, a model-independent prediction is the existence of $R$-axion which gets its mass exceptionally from gravitational instantons. 

Below we shall give a more detailed discussion of this scenario. We shall gauge the level of its uniqueness and outline some of its consequences. 
   
\section{Instantons}
Instantons play a crucial role in non-abelian gauge theories. Due to them, the theory possesses an infinite number 
of ground states corresponding to 
distinct super-selection sectors \cite{Callan:1976je, Jackiw:1976pf}. The vacua, $\ket{\theta}$, are usually labelled by a continuous periodic parameter $\theta$.  
The important point is that  $\theta$-vacua  are physically distinct. In particular, they differ by the strength 
of $CP$-violation. They also differ in energy.  
More importantly for the present discussion, 
the energy of the
 vacuum state $\ket{\theta}$ is 
 $\theta$-dependent \cite{Callan:1977gz} and $\theta=0$ is a global minimum \cite{Vafa:1984xg}. 
   
 Before discussing the analogous vacuum structure in 
 GR, let us briefly review the origin of 
 $\theta$-vacua in $SU(N_c)$ Yang-Mills (YM) gauge theory. The topology of the vacuum manifold is 
a $3$-sphere with non-trivial homotopy 
\begin{equation}
    \pi_3(SU(N_c)) = Z\,.
\end{equation}
 Due to this, classically,  the vacua are characterized 
 by a winding number $n$  corresponding to the fundamental group $\pi_3$. However, in quantum theory, there are transitions
  mediated by instantons that change $n$.   
 The  corresponding tunneling rate for $\Delta n =1$ is
\begin{equation}
    \mathcal{M}\sim e^{-S} \,,
\end{equation}
where $S = \frac{8\pi^2}{g^2}$ is the  Euclidean action of the instanton solution, with $g$ the running gauge coupling.  
The vacuum states that are the eigenstates of the transition operator have the form, 
\begin{equation}
    \ket{\theta}=\sum_n e^{i\theta n}\ket{n} \,.
\end{equation}
 In pure YM theory, one can choose an arbitrary
 $\theta$ as a valid vacuum.  
This choice is accounted for by the inclusion of 
the following term in the Lagrangian,
\begin{equation}
    \mathcal{L}_{\theta} \, = \, 
    \theta \frac{g^2}{16\pi^2}G\Tilde{G} \,, 
\end{equation}
where $G_{\mu\nu}$ is the gluon field strength matrix, 
and  $G\Tilde{G} \equiv {\rm tr} G_{\mu\nu} G_{\alpha\beta} 
\epsilon^{\mu\nu\alpha\beta}$. 

The vacuum energy  $E_{vac}$ is a periodic function of $\theta$
with the global minimum at $\theta =0$ \cite{Vafa:1984xg}.  
 In a dilute instanton gas approximation, it has a simple form 
 \begin{equation} \label{Evac}
   E_{vac} (\theta) \, \propto \, - \cos{\theta}.
\end{equation}
However, the precise form is unimportant for our discussions. 

The existence of physical $\theta$-vacua is 
linked with the topological susceptibility of the vacuum which is 
given by the following correlator evaluated at zero momentum,  
\begin{equation} 
    \mathrm{FT} \langle G\Tilde{G}(x)~G\Tilde{G}(0)\rangle_{p\rightarrow 0}\, \label{GG_cor} 
\end{equation}
where ${\rm FT}$ stands for the Fourier-transform and $p$ is a four-momentum. The $\theta$-vacua exist if the above correlator is a non-zero constant.  

The topological susceptibility of the vacuum offers an alternative description of  $\theta$-vacua in the language of a massless $3$-form \cite{Dvali:2005an}. In order to see this, let us consider the 
  Chern-Simons $3$-form, 
    \begin{equation} \label{CS}
   C_{\mu\nu\alpha} \equiv {\rm tr} (A_{[\mu}\partial_{\nu}A_{\alpha]} + \frac23 A_{[\mu}A_{\nu}A_{\alpha]}) \,.
   \end{equation}
 Here, $A_{\mu}$ is the rescaled gluon field matrix, $A_\mu\rightarrow gA_\mu$, which 
   under $SU(N_c)$ gauge redundancy transforms as 
   \begin{equation}\label{GaugeT}
   A_{\mu} \rightarrow U(x)A_{\mu} U^{\dagger}(x) 
+  U^{\dagger} \partial_{\mu} U     
   \end{equation}
with  $U(x) \equiv e^{-i\omega(x)^bT^b}$, where $T$-s are generators and $\omega$-s are transformation parameters. 
 The term $G\Tilde{G}$  represents a gauge 
 invariant field strength of $C$,  $G\Tilde{G} = \epsilon^{\alpha\beta\mu\nu} \partial_{\alpha} C_{\beta\mu\nu}$.   
A non-zero (\ref{GG_cor}) implies
 that a Chern-Simons $3$-form contains 
 a massless field  \cite{Luscher:1978rn}. This fully determines the vacuum structure of the theory, since 
  in EFT of a massless $3$-form, 
 the vacuum expectation value (VEV) of  $G\Tilde{G}$
 can take continuous values parameterized by an 
 integration constant $\theta$ \cite{Dvali:2005an}.  This is the physical meaning of  
 $\theta$-vacua in  the $3$-form language. 
 
In pure $SU(N_c)$, the correlator 
(\ref{GG_cor}) is non-zero due to instantons. 
The instanton calculus can be trusted as long as the 
action of the instanton is large, $S \gg 1$.    
This is guaranteed in the weak coupling regime. At the scale where the action becomes order one or smaller, 
 the instanton calculus ceases to make sense and the theory enters a strong coupling regime.   The corresponding 
 strong coupling scale,  usually referred to as the QCD scale,  marks the cutoff of EFT description in terms of gluons.
    
\section{The strong-CP problem}

In ordinary  QCD, the existence of $\theta$-vacua 
is the source of the famous strong-CP problem \cite{Callan:1976je, Jackiw:1976pf}. 
 The essence of it is that the $CP$-violating  
 $\theta$-parameter is experimentally constrained
  to $\theta  \lesssim 10^{-10}$ from the  
  measurements of the  electric dipole moment of neutron \cite{Abel:2020pzs}.  On the other hand, in pure 
$SU(N_c)$  this parameter is arbitrary. 
 In this light, the current phenomenological value seems inexplicably small. 
 
 Traditionally, this is viewed as the naturalness puzzle. However, the $S$-matrix formulation 
 of gravity promotes the strong $CP$ puzzle into a consistency issue,   demanding that  $\theta$ must be strictly unphysical \cite{Dvali:2022fdv, Dvali:2023llt}. 
 
In the standard approach, this is achieved by 
introduction of spin-$1/2$ fermion(s), $\psi$, transforming 
 under an anomalous  $U(1)$ chiral symmetry, 
\begin{equation} \label{PQ}
    \psi \rightarrow e^{i\gamma_5\alpha} \psi \,, 
\end{equation}
 where above is written in the Dirac basis. 
One can immediately anticipate that such a setup must render $\theta$
unphysical, provided the chiral symmetry is exact 
modulo the gauge anomaly. Indeed, the 
$\theta$-parameter shifts as 
$\theta \, \rightarrow \, \theta \, + \,2\alpha$ under the above transformation due to the anomaly.  Correspondingly, one can remove any would-be observable effect of $\theta$ by field redefinition.  

 This expectation is fully matched by the actual dynamics of the theory. 
The effect can be understood as the influence of the chiral fermions on instanton transitions. 
 The above shift is generated via the anomalous relation,
\begin{equation}
    \partial_{\mu} j_5^{\mu}=\frac{g^2}{8\pi^2}G\Tilde{G},
\end{equation}
where $j_5^{\mu}$ is the Noether's current of the chiral $U(1)$.  Integrated, the above equation gives the following  expression 
 for the charge,
\begin{equation}
    Q_5(t=\infty)\, -\, Q_5(t=-\infty)\, = \,2n,
\end{equation}
where $n$ is the index of the instanton. The last equation tells us that the change of the fermion chirality compensates for the transitions. Therefore, in the presence of chiral fermions the $\theta$ vacuum landscape ceases to exist.

In the language of instanton zero modes, a chiral fermion delivers a zero mode in the instanton background, which suppresses the transitions. 
 We can have the following two generic realizations of the chiral fermion scenario. 
 
  This first scenario called the Peccei-Quinn solution
  \cite{Peccei:1977ur, Peccei:1977hh},
  is realized if $\psi$ is massive. 
  This feature is shared by all realistic implementations (for a review, see, e.g., \cite{Kim:2008hd}).  
  In this case,  the chiral $U(1)$-symmetry 
 must be spontaneously broken by an additional complex scalar field
 $\Phi$ (elementary or composite) with the coupling 
 \begin{equation} \label{PQpsi}
   |\Phi| e^{-i \frac{a(x)}{f_a}} \bar{\psi}\psi \,, 
\end{equation}
where $f_a$ is the VEV of the scalar field modulus and 
$a$ is the phase degree of freedom called the axion 
\cite{Weinberg:1977ma, Wilczek:1977pj}. Although, the fermion is massive, the chiral $U(1)$-symmetry (\ref{PQ}) remains exact at the 
perturbative level.  The  transformation of 
 fermions is  compensated by the corresponding shift of the axion: 
  \begin{equation} \label{aShift}
    a(x) \, \rightarrow \, a(x)\, + \, 2\alpha\, f_a\,.
 \end{equation}   
   Via the chiral anomaly, the axion acquires the following coupling, 
\begin{equation}
    \mathcal{L} \, = \, \frac{g^2}{16\pi^2} \, \frac{a}{f_a} \, G\Tilde{G}\,,
\end{equation}
 thereby making $\theta$ dynamical. The 
 instanton-induced vacuum energy (\ref{Evac}) then translates 
 into the axion potential, relaxing $\theta$ 
 to zero. Simultaneously,  the same potential generates 
  the mass of the axion. 

  A somewhat different scenario emerges when the fermion 
  $\psi$ is massless. In this case, the
  chiral symmetry is spontaneously broken by the QCD condensate
  $\langle \bar{\psi} \psi \rangle \neq 0$.        
 The role of the axion is played by the phase 
 of the condensate.   For example, in standard QCD 
 this role would be performed by the $\eta'$-meson   
 provided one of the quarks would have a vanishing Yukawa coupling 
 with the Higgs.
 
  As shown in \cite{Dvali:2005an}, in the language of 
  topological susceptibility   \eqref{GG_cor},  both scenarios can be described as a $3$-form Higgs effects.  In this process, 
  the axion is ``eaten up'' by the  
  Chern-Simons $3$-form which becomes massive. 
  Correspondingly, the massless pole in the 
   topological susceptibility   \eqref{GG_cor} is removed and the correlator vanishes,   
   \begin{equation}
    \mathrm{FT} \langle G\Tilde{G}(x)~G\Tilde{G}(0)\rangle_{p\rightarrow 0}
    \propto \left. {\frac{p^2}{p^2-m^2}} \right\vert_{p\rightarrow 0} \,= \,0,
\end{equation}
where $m$ is the mass of the corresponding axion. 
In the Peccei-Quinn scenario, the axion is an elementary pseudo-scalar, 
whereas, in the case of a massless quark, it is represented by a composite $\eta'$-meson.

\section{Axion quality and the gauge axion.}\label{gauge_axion}

As already discussed, the $S$-matrix constraint on the vacuum landscape demands a complete elimination of the $\theta$-vacuum structure in each gauge sector of the theory \cite{Dvali:2022fdv, Dvali:2023llt}. This translates into the requirement of the 
exactness of the axion mechanism. In particular, in the case of a QCD axion,  introduced via a Peccei-Quinn or a massless quark mechanism, 
the global $U(1)$-symmetry must be broken exclusively 
by the QCD instanton effects.  Any other source of explicit breaking is strictly forbidden. 

This promotes the ordinary ``axion quality problem" into a consistency issue since any continuous deformation 
of the theory by explicit-breaking operators makes it inconsistent with gravity.    Thus, $U(1)$-breaking 
operators must vanish 
to all orders in the operator expansion \cite{Dvali:2022fdv, Dvali:2023llt}.

This requirement can in principle be adopted as a consistency prescription. However,  such a treatment would be rather unsettling.  Namely,  
it would be a mystery 
how the requirement of the gravitational consistency penetrates 
into the low-energy EFT.  

 The point is that the Peccei-Quinn mechanism 
 can be self-consistently implemented in the 
 limit  $M_{pl} \rightarrow  \infty$ while $f_a$ and all parameters of the Standard Model are kept finite. 
  Such a theory can be continuously deformed by $U(1)$-violating operators already at a renormalizable level.
No matter how mild,  this results in the appearance of a massless pole in the vacuum correlator (\ref{GG_cor}) and the corresponding re-emergence of the $\theta$-vacua. However, by consistency with 
 gravitational physics, this should not be possible. Since for any finite value of $M_{pl}$ such operators must 
 be strictly zero, by continuity, they must vanish also 
 for   $M_{pl} = \infty$. Although this would be fully understandable 
 for $1/M_{pl}$-suppressed operators induced by gravity, it is unclear why 
 the same should apply to arbitrary operators. 
 Thus, somehow already a non-gravitational theory must secretly know about the constraint coming from gravity.  
 
 While there is no {\it a priory}  argument against such
 miraculous cancellations, it is much more reasonable to expect that there exists a protecting mechanism already at the level of the low-energy EFT, regardless of the value 
of $M_{pl}$ \cite{Dvali:2022fdv, Dvali:2023llt}. An example of such built-in protection is provided by 
 a gauge formulation of the axion, proposed in \cite{Dvali:2005an}.
 In this theory, axion is an intrinsic part of the QCD-gauge redundancy without invoking any global symmetry. In this picture,  axion is introduced as a $2$-form  Kalb-Ramond field, $B_{\mu\nu}$. 
  The key point is that $B_{\mu\nu}$ is subjected to gauge redundancy  
  of  $SU(N_c)$-theory under which the gluon 
  matrix transforms as (\ref{GaugeT}).   

Under this redundancy, the gauge axion shifts as 
\begin{equation} \label{BQCD}
B \rightarrow B + \frac{1}{f_a}\Omega,
\end{equation}
where $\Omega_{\mu\nu} = {\rm tr} A_{[\mu}\partial_{\nu]}\omega$. At the same time, the Chern-Simons $3$-form (\ref{CS}) transforms as, 
\begin{equation} \label{BQCD1}
C_{\mu\nu\beta} \rightarrow C_{\mu\nu\beta} + \partial_{[\mu}\Omega_{\nu\beta]} \,. 
\end{equation}
 As a result, $B_{\mu\nu}$ can enter the Lagrangian exclusively through the following gauge invariant combination, 
 \begin{equation}\label{Cbar}
 C_{\mu\nu\beta} - f_a\partial_{[\mu}B_{\nu\beta]}\,.
 \end{equation}  
 In the lowest order, this adds to the YM Lagrangian 
 the term \cite{Dvali:2005an},  
  \begin{eqnarray} \label{TheA} 
 L \, &=& \, \frac{1}{f_a^2}(C \, - \, f_adB)^2 \,.
\end{eqnarray}
 That is, $B_{\mu\nu}$ acts as a St\"uckelberg component for 
 $C_{\mu\nu\beta}$ which makes the $3$-form massive. 
   This removes the massless pole in the   $3$-form propagator 
   and makes the topological susceptibility of the vacuum 
   zero, thereby eliminating $\theta$-vacua. 
   
Due to the gauge invariance, this formulation of axion is insensitive to continuous deformations of the theory by arbitrary operators  \cite{Dvali:2022fdv,Dvali:2005an,Sakhelashvili:2021eid}. Naturally, such axion is free of the quality problem. 

\section{Gravitational \texorpdfstring{$\theta$}{LG}-vacua} 

 We now wish to apply the  $S$-matrix criteria to gravitational 
 analogs of $\theta$-vacua.  For this, in the first place, we must ask whether 
 there exist instanton solutions in GR that give rise 
 to such vacua. 
    We shall assume that we are in an asymptotically Minkowski space and shall later verify the self-consistency 
    of this assumption.   
   We look for instanton solutions that 
   have topologies of three-spheres up to a discrete stabilizer
   and are Ricci flat with well-defined asymptotics.   

 The well-known solutions of the above type 
 are Eguchi-Hanson instantons \cite{Eguchi:1978xp,Eguchi:1978gw} which correspond to the following metric,
\begin{equation}
    ds^2=\left(1-\frac{a^4}{r^4}\right)^{-1}dr^2+r^2\left(\sigma_x^2+\sigma_y^2\right)+r^2\left(1-\frac{a^4}{r^4}\right)\sigma_z^2,
\end{equation}
where $a$ is the size of the instanton and $\sigma$-s are $SU(2)$ Cartan forms. For large distances, the metric is approaching $S^3/Z_2$, while the origin at $r=a$ is represented by $S^2$. There are two topological invariants associated with these instantons. One is the Euler characteristics $\chi$, which is defined by the Gauss-Bonnet topological term up to the boundary terms. Because of non-contractable $S^2$, it is equal to,  
\begin{eqnarray}\label{GB_chi}
\chi &\coloneqq & \frac{1}{8\pi^2}\int d^4x\sqrt{g}\left(R^2-4R_{\mu\nu}^2+R_{\mu\nu\alpha\beta}^2\right)+\mathrm{bound.~terms} \nonumber \\
&=& 2.
    \label{Euler}
\end{eqnarray}
The second is the gravitational Pontryagin index which is equal to, 
\begin{eqnarray}
\tau &\coloneqq & -\frac{1}{24\pi^2}\int d^4x \, R\Tilde{R} \nonumber \\
&=1.&
\label{tau_inv}
    \label{Euler1}
\end{eqnarray}
[$\Tilde{R}R \coloneqq \epsilon^{\mu\nu\alpha\beta} R_{\gamma \mu \nu}^{\kappa} R_{\kappa \alpha \beta}^{\gamma}$ is the Chern-Pontryagin scalar density]. Thus, similarly to the YM case, the above instanton is expected 
 to induce tunnelling transitions between different would-be vacuum states of the classical theory. However, there is a very important caveat that requires a proper interpretation. 
 
 Namely, in pure GR, the 
 above configurations have zero actions,
\begin{equation}
    S \, = \,0 \,.
\end{equation}
The existence of an instanton with zero 
 action implies that EFT, which we originally thought to 
 be weakly-coupled, in fact, is not. 
 That is, the quantum process mediated by the given instanton is maximally sensitive to the cutoff physics.
Universally, a zero action instanton puts 
 EFT out of the domain of validity. 
 
 In order to make sense 
 of the effect, the instanton action must be properly
 regularized by the additional terms in the gravitational action. 
  The regulator terms must be maximally ``harmless" in the sense that the perturbative infrared gravitational physics must be 
  insensitive to their presence. A unique invariant of this sort is the $\chi$-term given by \eqref{GB_chi},
which was omitted from our minimal construction.
   This term has to be added to the Euclidean action, 
   \begin{equation} \label{GB}
    \Delta S\, = \, c \, \frac{\chi}{2}\,,
\end{equation}
where $c$ is a positive dimensionless parameter. 

The above term is usually neglected since it is a boundary term and does not change the perturbative physics.
 Due to this property, even in perturbative renormalization of gravity, 
 the Gauss-Bonnet term is not taken into account \cite{Veltman:1975vx,tHooft:1974toh}. However, in our case, this term has a crucial role. It gives an extra contribution to the action of the instanton making it non-zero. In the new theory, the instanton action is proportional to the constant $c$ \footnote{The theory prohibits any solutions with negative Euler characteristics, they render the action to be negative. Also, the inclusion of the Gauss-Bonnet term enhances the instability of de Sitter space through the nucleation of black holes \cite{Parikh:2009js}. } and the vacuum transition rates are given by,
\begin{equation} \label{rateinC}
    \mathcal{M}\sim e^{-c}.
\end{equation}
The validity of EFT demands $c \gg 1$.   

 Although $c$ is dimensionless, it ``secretly"  encodes 
information about the cutoff scale $\Lambda_{gr}$,  
\begin{equation}
    c\sim\left(\frac{M_{pl}}{\Lambda_{gr}}\right)^2\,.\label{c_para}
\end{equation}
 This can be seen from the effective coupling 
 of  canonically-normalized gravitons in the 
 Lagrangian (\ref{GB}). In other words, $1/c$ has a meaning of an effective gravitational coupling at the scale $\Lambda_{gr}$. 
 The value of $\Lambda_{gr}$ is not a concern for the present paper, as long as it is around or below $M_{pl}$. 

  With this condition satisfied, 
  Eguchi-Hanson instanton becomes a fully trustable
  configuration mediating the vacuum transitions. 
  It is therefore expected to create a gravitational analog of 
  $\theta$-vacua by contributing to the correlator, 
   \begin{equation}
    \mathrm{FT}\langle\Tilde{R}R(x)~\Tilde{R}R(0)\rangle _{p\rightarrow 0}\neq 0\,. \label{Grav_cor} 
\end{equation}
Indeed, adding an extra term to the action,
\begin{equation}
    S =\frac{\theta}{24\pi^2}\int d^4x R\Tilde{R},\label{grav_theta}
\end{equation}
and taking the second derivative with respect to the $\theta$-angle confirms this. Like in the case of YM, the gravitational 
instanton generates different vacua, labelled by the parameter $\theta$. The energy of the ground state will depend on it.  Thus, starting 
from a ``naive" semi-classical Minkowski vacuum, we obtained a landscape of $\theta$-vacua.

 Let us summarize the steps we have taken so far. 
At the start of the discussion, we choose our vacuum to be Minkowski. This vacuum admits Eguchi-Hanson instantons. The validity of EFT demands the largeness 
of the instanton action, $S \gg 1$, which necessitates the addition of the Gauss-Bonnet terms (\ref{GB}) to the action. 
 The vacuum transitions generated by instantons result in the appearance of gravitational $\theta$-vacua  parameterized by 
the term \eqref{grav_theta}. 
  Thus, starting with the Minkowski vacuum, we ended up 
  with a landscape of non-degenerate $\theta$-vacua.  
 This situation is neither consistent nor compatible 
 with the $S$-matrix criterion of gravity.  
  Thus, gravity requires a mechanism that
 eliminates $\theta$-vacua. 
  The physicality of the term \eqref{grav_theta} and the necessity to make it unphysical shall be called {\it the gravity-$CP$ problem}.

\section{Seeking solutions to the gravity-\texorpdfstring{$CP$}{LG} problem}

Following in the same footsteps as in the case of YM, 
 we can try to eliminate the gravitational $\theta$-vacua
 by means of anomalous $U(1)$-symmetry.  
  Interestingly, this cannot be achieved by the spin-$1/2$ fermions.  
   
    Let us first consider the case of a massless spin-1/2 fermion
    $\psi$ transforming under the chiral symmetry 
    (\ref{PQ}).   It is well known that such a symmetry 
    suffers from a chiral gravitational anomaly \cite{Ganomaly1, Ganomaly2, Fujikawa:2004cx},             
\begin{equation} \label{gAnomaly}
    \partial_{\mu} j_5^{\mu} \propto R\Tilde{R}.
\end{equation}
From the first glance, the equation (\ref{gAnomaly}) leaves 
an impression that gravitational $\theta$ can be 
shifted away by the chiral transformation of the fermion 
(\ref{PQ}) thereby solving the gravity-$CP$ problem.  
 However, this is not the case. A careful computation of the index on the Eguchi-Hanson background \cite{Eguchi:1978gw} shows that
\begin{equation}
    Q_5(t=\infty)\, - \,Q_5(t=-\infty)\, = \,0.
\end{equation}
This expression tells us that a spin-$1/2$ fermion can not compensate the index of the instanton and the transitions between different topological vacua continue to happen even in the presence of a chiral fermion. 

The above observation is a particular case of a general index theorem \cite{Atiyah:1975jf} which states the absence of zero modes on the Eguchi-Hanson background for a spin-$1/2$ fermion. This is clear, since both the Ricci tensor and the Ricci scalar are zero on the background, leaving the Dirac operator unable to deliver zero modes.

Analogously fails the attempt of solving the gravitational-$CP$
 problem by a Peccei-Quinn-type mechanism 
 via the coupling (\ref{PQpsi}).   
 The presence of a Goldstone boson of spontaneously broken 
 $U(1)$ symmetry changes nothing since the intermediary between axion and the gravitational topological susceptibility is still a spin-$1/2$ fermion which has no index. 
 
 Thus, unlike the case of YM theories, the gravity-$CP$ problem cannot be solved by 
 spin-$1/2$ fermions.  We therefore conclude that pure GR, along with  
 its extensions via ordinary gauge theories, 
 exhibits the gravity-$CP$ problem. 
  
 The solution to this problem requires a rather profound extension of GR. 
 This brings us to our key point.
The fermion of the lowest spin that is capable of 
 eliminating the gravitational $\theta$-vacua is a chiral fermion 
 with spin$=3/2$.  Indeed, the index theorem shows a nontrivial index of helicity $\pm 3/2$ fermion in the Eguchi-Hanson background \cite{Eguchi:1978gw,Hanson:1978uv}. The chirality of the massless Rarita-Schwinger field in the above background is broken by two units \cite{Hanson:1978uv},
\begin{equation}
    I_{3/2}=-2\,.\label{ind_gravitino} 
\end{equation}
Correspondingly, coupling a massless helicity $\pm3/2$-field to GR solves the gravity-$CP$ problem. 

As it is well known \cite{Freedman:1976xh} (see e.g. \cite{Freedman:2012zz}),
a theory that includes a fermion of helicities $\pm 3/2$ coupled to gravity incorporates   
local supersymmetry, called supergravity.  
The spin-$3/2$ fermion in question, $\psi_{\mu}$, is gravitino. 
 The anomalous $U(1)$-symmetry that 
 renders the gravitational  $\theta$-vacua unphysical 
 is the $R$-symmetry under which gravitino transforms as, 
 \begin{equation} \label{Rsymmetry}
   \psi_{\mu} \rightarrow  {\rm e}^{i\alpha} \psi_{\mu}  \,.
 \end{equation}
 The  $R$-charge has an anomaly with respect to gravity \eqref{ind_gravitino}.  Therefore, performing $R$-rotation, we can arbitrarily redefine $\theta$ which indicates that it is 
 unphysical. The above index predicts two zero modes on the instanton background.  And indeed, there exist two zero modes \cite{Hawking:1978ghb, Konishi:1988mb}. \\

  This concludes the main message of the present paper: 
  \emph{GR suffers from the  gravity-$CP$ problem.
  The minimal consistent theory that eliminates this problem is supergravity}. \\

\section{Gravitational \texorpdfstring{$R$}{LG}-axion and SUSY breaking}

 The  existence of zero modes in the Eguchi-Hanson background generates a corresponding 't Hooft vertex 
for gravitino,  
      \begin{equation} \label{Gthooft} 
        \frac{W_{3/2}^*}{M_{pl}^2} \, \bar{\psi}^{\mu}\sigma_{\mu\nu} \psi^{\nu}\,,  
   \end{equation}
 where we have introduced a parameter $W_{3/2}$ describing its strength.
 Of course, it incorporates the exponential suppression by the 
 instanton action (\ref{rateinC}).  
  
The quantity $W_{3/2}$ can be interpreted as
a dynamically generated superpotential that breaks the 
anomalous $R$-symmetry via the gravitational instantons.  This contribution does not break supersymmetry but
it would lower the vacuum to AdS. 

   In order to keep the theory in Minkowski vacuum, 
a positive contribution to vacuum energy must 
appear that can balance the negative contribution from 
the superpotential.  
Such a contribution will necessarily break SUSY. Simultaneously, by consistency, the theory must provide a pseudo-scalar degree of freedom that eliminates the massless pole in the correlator (\ref{Grav_cor}) \cite{Dvali:2005an}.  

From the required quantum numbers and anomaly-matching, it is obvious that 
the corresponding pseudo-scalar must be a pseudo-Goldstone boson of spontaneously broken $R$-symmetry that gets its mass exclusively from the topological susceptibility of gravity.  We shall refer to this degree of freedom as $R$-axion and denote it by $a_R$. We are thus led to the predictions of the spontaneous SUSY-breaking and the accompanying $R$-axion.  

 The spontaneously broken $R$-symmetry must be exact, modulo the Eguchi-Hanson instantons. The remaining question is the identity of a sector that 
  is responsible for spontaneous  breakings of 
 SUSY and of $R$-symmetry. 
 
 One option that shall be briefly considered later is that 
both tasks are accomplished by a gravitino composite
multiplet.  Putting this option aside, here we shall 
demonstrate how this can be achieved by 
a minimal extension of the theory within weakly coupled 
EFT.  

The purpose of the present paper is far from building a complete phenomenologically viable model. However, we shall construct a consistent model that captures 
the required features. The superpotential generated by the gravitational 't Hooft vertex (\ref{Gthooft}), contributes the negative cosmological term,  $\propto  -3|W_{3/2}|^2/M_{pl}^2$. 

In order to break supersymmetry and to uplift the vacuum of the theory to Minkowski, it is sufficient to add a single chiral superfield $\hat{X}$, with the components 
  $X, \chi, F_X$, that enters linearly in the superpotential,  
\begin{equation} \label{P-W}
  W \, = \,   \hat{X}\Lambda_{X}^2  +  W_{3/2} \,.
\end{equation} 
 Here $\Lambda_{X}$ is a scale, which we take
 to be safely below the Planck mass  \footnote{Notice that such a linear superpotential can be generated dynamically, using the construction given in \cite{Dimopoulos:1997ww, Dimopoulos:1997je}.}.  This is because it has to match the superpotential generated by the gravitational 
 't Hooft vertex $W_{3/2}$ which is expected to be exponentially suppressed relative to the Planck scale. 
 
 We thus effectively end up 
with the  Pol\'onyi model  \cite{Polonyi} with the difference that the constant term in the superpotential is dynamically generated by instantons \cite{Konishi:1988mb, Konishi:1989em}. 
The $R$-charge of the $X$-superfield is equal to the 
$R$-charge of gravitino.  For simplicity we assume 
the minimal K\"ahler function for  $X$, 
$K = \hat{X}^{\dagger}\hat{X}$.  The minimization of the scalar potential is identical to the original  Pol\'onyi case.

Performing minimization and setting the vacuum energy to zero we obtain the 
  following VEV for $X$, 
  $X_0 \, = \, \pm M_{pl} (\sqrt{3} + 1)$.  At the same time, the parameters must be tuned to satisfy
  $W_{3/2} \, = \,\mp \, \Lambda_X^2 M_{pl} (\sqrt{3} + 2)$. In this way, the theory has a Minkowski vacuum  in which the SUSY-breaking 
$F_X$-term and the gravitino 
mass satisfy $|F_X| \, = \, \sqrt{3} \, \Lambda_X$ and 
 $m_{3/2} \, = \, W/M_{pl}^2 \, = \,  \Lambda_X^2/M_{pl} $
 respectively. Notice that the gravitino mass $m_{3/2}$ is much smaller 
  then the scale of gravitational 't Hooft vertex $\Lambda_{3/2} 
  \, = \, (W_{3/2})^{1/3}$: 
  \begin{equation} \label{mandCond}
  \frac{m_{3/2}}{\Lambda_{3/2}} \sim \left (\frac{\Lambda_{X}}{M_{pl}} \right )^{\frac{4}{3}}  \ll 1\,.
 \end{equation}     
 
Since the VEV of $X$ breaks $R$-symmetry spontaneously,  the $R$-axion, $a_R$, is the phase of $X$. The decay constant of this axion is around the Planck scale whereas its mass is of order $m_{3/2}$. The situation is similar to the story with hidden axion in QCD, in which the quark masses are generated from the Yukawa coupling (\ref{PQpsi}) with the Peccei-Quinn field.
      
As it is standard for a super-Higgs effect, 
the gravitino mass comes from the gravitino bilinear term as well as from the mixing with Goldstino, the role of which is played by the chiral fermion  component, 
$\chi$, of the $\hat{X}$-superfield, 
    \begin{equation}
        \frac{W^*}{M_{pl}^2} \, \bar{\psi}^{\mu}\sigma_{\mu\nu} \psi^{\nu}\, + \,   \frac{F_X^*}{M_{pl}} \, \bar{\chi} \gamma_{\mu} \psi^{\mu}\,. 
   \end{equation}
Notice that $F_X$-term carries the zero $R$-charge so that the $R$-charges of Goldstino and gravitino are equal. At the same time, the $R$-charge of gravitino bilinear is compensated by the $R$-charge of the superpotential. 
   
We emphasize that, although the $R$-axion is an elementary phase degree of freedom coming from $X$,  the communication with the gravitational topological susceptibility takes place via the spin-$3/2$ gravitino.  That is, the existence of  $\pm 3/2$ helicity zero modes in the instanton background is crucial for eliminating the gravitational $\theta$-vacua. 

The existence of $R$-axion 
with the scale $\sim M_{pl}$ and mass $\sim m_{3/2}$, 
is a model-independent outcome of the gravity-$\theta$ problem. 
 This particle can have interesting phenomenological implications. 
 In particular, it can constitute the dark matter in the current Universe.

\section{Constraint on chiral charges of
spin-\texorpdfstring{$1/2$}{LG} fermions}

The present picture imposes an important constraint on the chiral $U(1)$-charges of the spin-$1/2$ fermions.  Consistency with the
index computation \cite{Eguchi:1978gw} requires the contribution to the chiral gravitational anomaly to originate exclusively from the spin-$3/2$ gravitino.
 This implies that the contribution to the chiral gravitational anomaly from the rest of spin-$1/2$ fermions must be exactly zero. 
 
Indeed, in the opposite case, there would be a mismatch 
 between the gravitational chiral anomaly 
 (\ref{gAnomaly}) and the index theorem\footnote{One must point out that the perturbative gravitational chiral anomaly, unlike its Yang-Mills counterpart, may not be one-loop exact.  However, we assume that even if non-zero, the higher order corrections are subdominant and the
 relation (\ref{gAnomaly}) can be trusted on the Minkowski vacuum. }. 
 In particular, it would be possible to 
 rotate away the gravitational $\theta$-term (\ref{grav_theta}) 
 via a chiral transformation of some spin-$1/2$ fermions 
 without the need for a spin-$3/2$.   
     This would be a clear inconsistency, since such fermions 
    possess no zero modes in the background 
    of Eguchi-Hanson instanton, and thereby, are unable 
    to get rid of the topological susceptibility of 
    the gravitational vacuum (\ref{Grav_cor}). 
    Thus, we are led to the following powerful conclusion: \\ 
    
   \noindent \emph{
1) $R$-symmetry must be explicitly broken exclusively by the gravitino anomaly;
2) the contribution to the chiral gravitational anomaly from a $U(1)$-symmetry acting on a spin-$1/2$ fermion must vanish, or else, the symmetry must be explicitly broken by other sources. } \\
    
 In other words, a consistent supergravity theory 
 forbids a spin-$1/2$ fermion with a global $U(1)$ charge broken exclusively via the gravitational chiral anomaly (\ref{gAnomaly}).  In the opposite case, the theory must respond by invalidating the description. 
 
   Let us illustrate this on the previously discussed Polonyi model.  Naively, it looks very easy to circumvent 
   the above constraint by coupling  the  Pol\'onyi  superfield $\hat{X}$  to  a set of chiral superfields $\hat{Y}_j$, where $j = 1,2,...,N_F$ is a flavor index. The superpotential
   can be chosen in the following form,   
   \begin{equation} \label{NewW}
  W \, = \,   \hat{X}\Lambda_{X}^2   -  g  \hat{X} \hat{Y}_j^2
  +  W_{3/2} \,,
\end{equation} 
where $g$ is a coupling, which for simplicity we took to be
flavor-blind.  
The fermion components of  $\hat{Y}_j$-superfields
carry one unit of the $R$-charge. It is clear that the contribution from these fermions
into the gravitational anomaly (\ref{gAnomaly}) of 
$R$-symmetry is non-zero and is proportional to 
$N_F$.   At the same time, such fermions do not 
contribute any additional zero modes into 
Eguchi-Hanson instantons.  
It thus appears that a conflict between the 
index theorem and the gravitational anomaly is imminent. 

 However, this conclusion is premature:  the 
coupling to $\hat{Y}_j$-s restores 
supersymmetry by destabilizing 
the would-be  Minkowski vacuum and creating 
a new supersymmetric  AdS vacuum with 
$g X Y_j^2 \simeq \Lambda_{Y}^2$, where 
$Y_j$-s are scalar components. In order to save the Minkowski vacuum, we must maintain broken supersymmetry by 
 preventing  $Y_j$-s from cancelling the 
 $F_X$-term.   This requires  an additional set of 
 superfields $\hat{\bar{Y}}_j$ with the couplings, 
    \begin{equation} \label{NewW1}
  W \, = \,   \hat{X}\Lambda_{X}^2   -  g  \hat{X} \hat{Y}_j^2
   +  M  \hat{\bar{Y}}_j \hat{Y}_j 
  +  W_{3/2} \,,
\end{equation} 
 where $M$ is  a
 mass parameter, which we took flavor-independent.    
 
In this way, the Pol\'onyi model 
is resolved as  the O'Raifeartaigh model \cite{OR-model}
which is known to break supersymmetry spontaneously. 
In particular, for $|M|^2 \, \gg \,  |2g  \Lambda_{X}^2|$, 
 we have $Y_j =0$ and SUSY is spontaneously broken by 
the $F$-term of the $\hat{X}$-superfield, 
$F_X \simeq \Lambda_{X}^2$. 

  In this regime, we can integrate the heavy superfields out 
  and the effective superpotential becomes the one of 
  Pol\'onyi (\ref{P-W}) with all the previous consequences intact. The important thing is that the 
  contributions  to the gravitational  $R$-anomaly from 
  $\hat{Y}_j$ and $\hat{\bar{Y}}_j$ superfields 
  cancel out.  
   We thus see that the existence of a consistent 
   Minkowski vacuum forces upon us the 
   spin-$1/2$ content with 
   a zero net contribution to the gravitational anomaly. 
    This property must be fulfilled in general.  
   
We must note that when introducing the superfields 
 charged under  YM gauge symmetries, in general, a new set of 
 anomalies appears.  Some of the chiral symmetries of 
 spin-$1/2$ fermions will be explicitly broken 
 by the respective gauge anomalies. This effect will get rid of the $\theta$-vacua in the corresponding gauge sectors. 
 Of course, the symmetries that are explicitly-broken 
 by other anomalies, are useless for the elimination of the gravitational  $\theta$-term\footnote{However, such fermions can deposit zero modes in mixed YM-gravity instantons \cite{Chen:2021jcb} and can eliminate one combination of gravitational and YM $\theta$ parameters, but never both.  Naturally, in each case, the anomaly and the 
index must consistently match. }. 
 For this, the presence of spin-$3/2$ is a necessity.      
      
The impossibility of rotating the gravitational $\theta$-term 
 by a chiral transformation of a spin-$1/2$ fermion has 
 important consequences.  For example, it tells us that 
 the neutrino masses cannot originate 
 from a topological susceptibility of gravity  \cite{Dvali:2016uhn}
 coming from Eguchi-Hanson instantons. Also, this provides a useful constraint on the 
  model-building for the neutrino mass in general. 
  A chiral symmetry that is spontaneously broken by 
 the mass of a neutrino, must either be free of the gravitational anomaly or be explicitly broken by other sources.

\section{Gravitino condensate}
 
Here we briefly touch on the possibility of a scenario 
 in which the $R$-axion represents (or has a substantial contribution) from 
 a gravitino composite. 
It has been argued \cite{Hawking:1978ghb, Konishi:1988mb,Konishi:1989em} that a bilinear condensate of gravitino,
\begin{equation} \label{gravitinoVEV}
\langle \bar{\psi}^{\mu}\sigma_{\mu\nu} \psi^{\nu} \rangle \neq 0 \,,
\end{equation}
is formed. The condensate must be accompanied by the appearance of a  composite multiplet, which consists of 
a pseudoscalar, a dilaton and a dilatino. The above is in agreement with the index \eqref{ind_gravitino}. The condensate includes two fermions and violates the $R$ charge by two units. In this scenario, the composite pseudoscalar, 
  $\eta_R$,  would get the 
 mass from the gravitational topological susceptibility 
 of the vacuum.
 
 This effect is similar to the 't Hooft mechanism of generation of the mass of 
  $\eta'$-meson from the instantons of QCD \cite{tHooft:1976rip, tHooft:1976snw}. 
 If one of the quarks were massless, the  $\eta'$-meson would 
 play the role of the axion and would fully cancel the $\theta$-parameter \cite{Dvali:2005an}.
 The explicit breaking of chiral symmetry by non-zero quark masses prevents the realization of such a scenario.  In QCD, the lattice results appear not to favor a massless quark scenario (for recent updates, see \cite{Alexandrou:2020bkd})

  From the gravitino condensate scenario
  \cite{Hawking:1978ghb, Konishi:1988mb,Konishi:1989em}, we expect the analogous picture.  The instantons form an order parameter (\ref{gravitinoVEV}) out of gravitino, which breaks the $R$-symmetry spontaneously. The corresponding Goldstone $\eta_R$ is ``eaten-up" by the $\theta$-vacuum, rendering it massive and simultaneously lifting the supper-selection.
  
  However, such a scenario leaves the following two questions open. 
  The first one is the origin of  SUSY-breaking which 
  is necessary for uplifting the vacuum to Minkowski. 
   The second question is about the form of EFT
  in between the compositeness scale of $\eta_R$ and $m_{3/2}$. 
  
A priory, one cannot exclude that both 
effects,  the formation of gravitino condensate as well as 
SUSY breaking by its composite multiplet,  
takes place at a scale close to the cutoff of the theory,
$\Lambda_{gr} \sim M_{pl}$. Such a scenario would eliminate 
the question about the low-energy EFT, since 
 $m_{3/2}$ would also be around the cutoff. However, for this scenario to work, the instanton action must be close to one.  This puts EFT at the boundary of the domain of calculability. 
At the same time, not much SUSY-dynamics would be left in EFT below $\Lambda_{gr}$\footnote{We remark that the understanding of $\eta_R$-mass via a reasoning analogous 
to the one of Witten-Veneziano for $\eta'$ in QCD with large number of colors \cite{Witten:1979vv,Veneziano:1979ec} requires a low cutoff,
$\Lambda_{gr}\, \ll \,  M_{pl}$.  
 In gravity the analog of a ``color" 
is played by the number of particle species, $N_{sp}$, 
 which is known to lower the non-perturbative cutoff imposed by black hole physics to the ``species scale"  $\Lambda_{gr} = M_{pl}/\sqrt{N_{sp}}$
 \cite{Dvali:2007hz,Dvali:2007wp}. }.

 The question of SUSY-breaking by a minimal theory requires further investigation.  Of course,  breaking of SUSY can be achieved at the expense 
 of additional multiplets. For example,  Konishi showed that the introduction of 
a single chiral superfield due to anomaly breaks supersymmetry 
\cite{Konishi:1988mb, Konishi:1989em}. And, of course, we can always resort to Polonyi or O'Raifeartaigh models, as we did in the previous sections. The question is whether one can get away with pure supergravity without additional chiral superfields \cite{susynew}.    
   
\section{\texorpdfstring{$R$}{LG}-axion quality and the gravitational gauge axion.}

Here we would like to address the following two questions: \\
     
   {\it 1)}  What can one say about the quality problem for gravitino  
  $R$-axion?  
  \\
     
 {\it 2)} Can a gauge axion-type mechanism for gravity \cite{Dvali:2005an} 
 serve as an alternative to a spin-$3/2$ fermion? \\
    
The answer to the first question is that, in contrast to the Peccei-Quinn axion,  in the case of spin-$3/2$ fermion the gravitational quality problem cannot be defined.  The reason is that gravitino and graviton cannot be decoupled separately from low-energy EFT. That is, unlike the Peccei-Quinn case, there exists no consistent limit 
in which gravity decouples while the gravitational instantons 
and gravitino remain interacting. Therefore, there exists no analog of the  Peccei-Quinn type gravitational quality problem. 
   
The answer to the second question is less straightforward. At first glance, a formulation analogous to the gauge axion \cite{Dvali:2005an},
reviewed in Sec. \ref{gauge_axion}, can be adopted in the present case. 
For this, one could introduce a $B_{\mu\nu}$ field which acts as St\"uckelberg  for the gravitational Chern-Simons 
  $3$-form: 
  \begin{equation} \label{CSG}
   C_{\mu\nu\alpha}^{(g)} \equiv {\rm tr} (\Gamma_{[\mu}\partial_{\nu}\Gamma_{\alpha]} + {\frac23}\Gamma_{[\mu}\Gamma_{\nu}\Gamma_{\alpha]}) \,,
   \end{equation}
   where $\Gamma_{\alpha}$ is the gauge-connection. 
 One then expects that this will remove the massless 
pole in the topological susceptibility of the gravitational vacuum, 
rendering $\theta$ unphysical. 

If a consistent realization of the above scenario is possible, it would serve as an alternative to supersymmetry for the elimination of gravitational $\theta$-vacua. However,  the validity 
of gravitational gauge axion requires an additional
comprehensive scrutiny against potential issues such as gauge anomalies \footnote{ For example, the analysis of anomalies for a minimally coupled 
Kalb-Ramond field can be found in \cite{Duff:1980qv}. Application of these results to the present case, in 
 which $B_{\mu\nu}$ exhibits a higher gauge redundancy of the 
 St\"uckelberg field, requires a special study.} which will be given elsewhere.

\section{Discussions}

The purpose of the present paper was to point out that  
certain very general constraints on the vacuum landscape, 
when applied to the topological structure of the gravitational vacuum, hint towards supersymmetry. Our input requirement is the absence of 
   $\theta$-vacuum structure in any sector of the theory, 
  including gravity \cite{Dvali:2018dce, Dvali:2022fdv, Dvali:2023llt}.
  This is justified by the $S$-matrix formulation of gravity. However, the reader can equally accept it as our starting point. 

This condition, applied to ordinary YM-type $\theta$-vacua, necessitates the existence of an axion-like degree 
of freedom that eliminates the massless pole in 
the vacuum correlator (\ref{GG_cor}). Within EFT this can be achieved in two ways. 
 The standard way is via an anomalous global Peccei-Quinn symmetry
 \cite{Peccei:1977hh}. 
 In all its incarnations, this requires the presence of some
  spin-$1/2$ fermions which deposit the fermionic zero modes in the corresponding instanton background. As noticed in \cite{Dvali:2005an}, in this sense, there is 
 no distinction between the Peccei-Quinn model with an elementary 
 scalar axion and a case of a massless chiral quark. 
 In the latter case, the axion is composite, with its role played 
 by the $\eta'$-meson.
 
 Alternatively, the axion can be introduced as a $2$-form Kalb-Ramond gauge degree of freedom without a reference to a global symmetry \cite{Dvali:2005an}. 

    In the present paper, we apply the requirement of absence of $\theta$-vacua to gravity. 
   First,  we argue that the consistency of EFT 
   demands the inclusion of the Gauss-Bonnet term. This term makes the action of 
    Eguchi-Hanson instantons non-zero.  Without 
    Gauss-Bonnet term this action vanishes, which
    signals that EFT is outside of the domain of its validity.
    
    With the Gauss-Bonnet term included, the finite-action instantons create a
   would-be $\theta$-vacuum which must be removed by 
   the consistency of our starting requirement. However, unlike the case of YM, this cannot be achieved 
    by a Peccei-Quinn mechanism with 
   a spin-$1/2$ fermion. The reason is that such fermions do not 
   generate zero modes in the background of Eguchi-Hanson instanton. 
   Due to this,  spin-$1/2$ fermions
   are powerless to eliminate the gravitational $\theta$-term. 

A fermion that is capable of performing such a task, is 
the one with spin$=3/2$.  The presence of such a fermion is a clear 
sign of supersymmetry.  Thus, the minimal fermionic structure compatible 
with our constraint is provided by supergravity. The way gravitino kills the gravitational $\theta$-term 
 is very similar to how a chiral quark performs the analogous task in 
 YM.
 
 However, in order to maintain a Minkowski vacuum, such a scenario necessitates 
 the spontaneous breaking of SUSY. The reason is that the contribution 
 from the gravitational 't Hooft 
 vertex in the superpotential provides a negative contribution to vacuum 
 energy.  For uplifting the theory ``back" to the Minkowski vacuum, this contribution 
 must be balanced by the positive contribution coming from 
 the spontaneous breaking of supersymmetry. Thus we arrive at a consistency requirement of simultaneous spontaneous breakings 
 of SUSY and of $R$-symmetry.   
 In addition, $R$-symmetry must be explicitly broken exclusively by the gravitational anomaly of gravitino, with zero contribution from 
 other sources, including anomalies of spin-$1/2$ fermions. 
 
Our model-independent prediction is the existence of $R$-axion, 
with a decay constant $M_{pl}$ and the mass $\sim m_{3/2}$ originating 
solely from the gravitational $\theta$-vacuum. This particle can have interesting implications, such as being a viable dark matter candidate. 

An important question is the origin of SUSY-breaking. 
In the present paper, we have limited ourselves to the simplest 
existence-proof realizations of  Pol\'onyi and O'Raifeartaigh models. Interestingly, it has been argued \cite{Hawking:1978ghb, Konishi:1988mb,Konishi:1989em} that instantons generate gravitino condensate 
(\ref{gravitinoVEV}) and form a corresponding composite multiplet. 
In this context,  a natural question would be whether such composites could
contribute to SUSY-breaking. We postpone the detailed study of this question 
to future work \cite{susynew}.
 
The necessity of spin-$3/2$ fermion for elimination 
of gravitational $\theta$-vacuum imposes an important constraint 
on the chiral charges of spin-$1/2$ fermions. 
Since such fermions possess no zero modes in the background 
of Eguchi-Hanson instanton, their contribution into 
the chiral gravitational anomaly (\ref{gAnomaly}) must vanish, 
or else, the corresponding chiral symmetry must be explicitly 
broken by external sources, such as, e.g., the YM gauge anomaly. This fact has important implications for the possible origin 
 of the neutrino mass from gravitational anomaly \cite{Dvali:2016uhn}.
 In particular, such masses cannot be generated from Eguchi-Hanson instantons.  
  
 The question that is left open in the present discussion is whether 
 instead of supersymmetry one can resort to a gauge axion version
 of gravity \cite{Dvali:2005an}. The consistency of such a mechanism requires an additional study. Even if gravitino is not the unique mechanism of eliminating the gravitational  $\theta$-vacua, it is certainly the one based on a fermion degree of freedom. This provides a strong motivation for supersymmetry at the level of embedding the Standard Model in the theory of gravity. \\

\section*{Acknowledgments} 

We would like to thank Lasha Berezhiani for the discussions and for pointing out to us Ref. \cite{Parikh:2009js}. AK and OS would also like to thank MPI-Munich for the hospitality extended during their visits when that part of this work was completed. 

The work of GD was supported in part by the Humboldt Foundation under the Humboldt Professorship Award, by the European Research Council Gravities Horizon Grant AO number: 850 173-6, by the Deutsche Forschungsgemeinschaft (DFG, German Research Foundation) under Germany’s Excellence Strategy - EXC-2111 - 390814868, Germany’s Excellence Strategy under Excellence Cluster Origins 
EXC 2094 – 390783311. The work of AK and OS was partially supported by the Australian Research Council under the Discovery Projects grants DP210101636 and DP220101721.     \\

\noindent {\bf Disclaimer:} Funded by the European Union. Views and opinions expressed are however those of the authors only and do not necessarily reflect those of the European Union or European Research Council. Neither the European Union nor the granting authority can be held responsible for them.\\

\end{document}